
\input phyzzx

\REF\HL {J.A. Harvey and J.T. Liu, {\sl Phys. Lett.} {\bf 268} (1991) 40.}
\REF\GKLTT {G.W. Gibbons, D. Kastor, L.A.J. London, P.K. Townsend and J.
Traschen, {\sl Nucl. Phys.} {\bf B416} (1994) 850.}
\REF\GHL{J.P. Gauntlett, J.A. Harvey, and J.T. Liu {\sl Nucl. Phys. }{\bf B409}
(1993) 363.}
\REF\Kh{R.R. Khuri, {\sl Nucl. Phys.} {\bf B387} (1992) 315.}
\REF\NT {H. Nicolai and P.K. Townsend, {\sl Phys. Lett.} {\bf 98B} (1981) 257.}
\REF\GST{M. G{\"u}naydin, G. Sierra and P.K. Townsend, Nucl. Phys. {\bf B242}
(1984) 244; {\sl Phys. Lett.} {\bf 144B} (1984) 41; {\sl Phys. Rev. Lett.} {\bf
53}
(1984) 322; {\sl Nucl. Phys.} {\bf B253} (1985) 573.}
\REF\M{K.E. M{\" u}ller, {\sl Phys. Lett.} {\bf B177} (1986) 389.}
\REF\GR {G.W. Gibbons and P.J. Ruback {\sl Phys. Rev. Lett.} {\bf 57}
(1986) 1492.}
\REF\Ru{P. J. Ruback {\sl Commun. Math. Phys.} {\bf 107}(1986) 93.}
\REF\Sh{K. Shiraishi {\sl Nucl. Phys.} {\bf 402} (1993) 399.}
\REF\Sor{R. Sorkin {\sl Phys. Rev. Lett.} {\bf 51} (1983) 87. }
\REF\GP{D. Gross and M. J. Perry {\sl Nucl. Phys.} {\bf B226} (1983) 29.}
\REF\Sen {A. Sen, {\sl Phys. Lett.} {\bf B329} (1994) 217.}
\REF\HP{S.W. Hawking and C.N. Pope, {\sl Nucl. Phys.} {\bf B146} (1978) 381.}
\REF\GM {G. W. Gibbons and N. Manton {\sl Nucl. Phys.} {\bf B274} (1986)
183.}
\REF\SM {B. Schroers and N. Manton {\sl Ann. Phys. (NY)} {\bf 225} (1993)
290.}


\font\mybb=msbm10 at 12pt
\def\bb#1{\hbox{\mybb#1}}
\def\R {\bb{R}}

\def\square{\kern1pt\vbox{\hrule height 0.5pt\hbox
{\vrule width 0.5pt\hskip 2pt
\vbox{\vskip 4pt}\hskip 2pt\vrule width 0.5pt}\hrule height
0.5pt}\kern1pt}


\Pubnum{ \vbox{ \hbox{R/95/22} } }
\pubtype{}
\date{June, 1995}

\titlepage

\title{Anti-gravitating BPS monopoles and dyons}

\author{G.W. Gibbons and P.K. Townsend}
\address{DAMTP, University of Cambridge,
\break
Silver St., Cambridge, U.K.}
\vskip 0.5 cm

\abstract{ We show that the exact static, i.e. `anti-gravitating', magnetic
multi monopole solutions of the Einstein/Maxwell/dilaton-YM/Higgs equations
found by Kastor, London, Traschen, and the authors, for arbitrary non-zero
dilaton coupling constant $a$, are equivalent to the string theory BPS magnetic
monopole solutions of Harvey and Liu when $a=\sqrt{3}$. For this value of $a$,
the monopole solutions also solve the equations of five-dimensional
supergravity/YM theory. We also discuss some features of the dyon solutions
obtained by boosting in the fifth dimension and some features of the moduli
space
of anti-gravitating multi-monopoles.}

\endpage

\pagenumber=2

It has been known for some time that certain non-abelian Yang-Mills/Higgs
theories in
flat spacetime admit multi-monopole solutions in which the magnetic repulsion
is
balanced by the attractive forces due to Higgs exchange. More recently it has
been
shown that this equilibrium continues to be possible in the presence of
additional
attractive forces due to  gravitation and a massless scalar field [\HL,\GKLTT].
In
[\GKLTT] this result was obtained directly in four dimensions by the inclusion
of an
additional abelian vector potential, $A_\mu$, having a non-renormalizable
coupling to the Yang-Mills magnetic charge density. Remarkably, it is then
possible
to find exact analytic solutions for the metric, dilaton and abelian vector
fields
entirely in terms of solutions of the {\sl flat space} Bogomol'nyi equations in
the
Yang-Mills/Higgs sector. These results were shown to hold for all
non-zero values of the `dilaton coupling constant', $a$, defined by the
coupling of
$\sigma$ to the Maxwell field strength, $F_{\mu\nu}$, provided that the scalar
field, $\sigma$, has particular couplings to the Yang-Mills gauge potential,
${\bf
B}_\mu$ through its field strength tensor ${\bf G}_{\mu\nu}$, and to the Higgs
field
${\bf
\Phi}$. These interaction terms might appear to be artificial but they are
precisely
those required by local supersymmetry (at least for certain values of $a$) and
therefore arise naturally in supergravity and superstring theories. The action
of
[\GKLTT] that was shown to admit these static self-gravitating solitons is
$$
\eqalign{
S= &{1\over4}\int d^4x\,\Big\{ \sqrt{-g}[ R - 2 ( \partial \sigma )^2 - e^{-2a
\sigma }F ^2 - e^{- {{(1-a^2)} \over a }\sigma }|{\bf G}|^2 - 2 e^{ {{1+a^2}
\over
a}\sigma } |{\cal D} {\bf \Phi}|^2 ] \cr
&\qquad\qquad  -2\sqrt {1+a^2}\, A_\mu\, \epsilon ^{\mu \nu \lambda \rho }
{\bf G}_{\nu\lambda}\cdot{\cal D}_\rho{\bf \Phi}\Big\} }
\eqn\aone
$$
where ${\cal D}$ is the YM covariant derivative, and we have set $4\pi G$ and
the
Yang-Mills (YM) coupling constant to unity. The dilaton coupling constant $a$
is
related to the constant $b$ used in [\GKLTT] by $a=-b$. We may choose $a\ge0$
without loss of generality, and we shall assume this in what follows.

The spacetime metric,
Maxwell one-form, and scalar field in the self-gravitating monopole solution of
[\GKLTT] have the form
$$
\eqalign{
ds^2 &= - U^{ {-2} \over {1+a^2}} dt^2 + U^{ {2} \over {1+a^2}}  d {\bf
x}^2\cr A
&= {{dt} \over U}  { 1 \over {\sqrt{1+a^2}} }\cr
\sigma
 &= -{a \over {1 + a^2}}\ln U \ .}
\eqn\atwo
$$
The function $U$ satisfies
$$
\nabla ^2 U = - (1+a^2) \sum_{i=1}^3| {\cal D}_i {\bf \Phi} |^2\ ,
\eqn\athree
$$
where ${\bf \Phi}$ is a solution of the {\sl flat space} Bogomol'nyi
equations:
$$
{\bf G}_{ij}= \delta_{il}\delta_{jm}\varepsilon^{lmk}{\cal D}_k {\bf \Phi}\ .
\eqn\afour
$$
Since the Bogomol'nyi equations imply that
$$
2 \sum_{i=1}^3 |{\cal D}_i {\bf \Phi}|^2 = \nabla ^2 \big(|{\bf \Phi}|^2\big)\
,
\eqn\afive
$$
and we require that $U\rightarrow 1$ at spatial infinity, we have that
$$
U = 1 + {1\over 2} (1+a^2) \big(\eta^2 - |{\bf \Phi}|^2\big)\ ,
\eqn\asix
$$
where $\eta$ is the value at infinity of $\sqrt{|{\bf \Phi}|^2}$,
so the solution is entirely, and explicitly, determined in terms of ${\bf
\Phi}$.
For example, for the $SO(3)$ BPS monopole we have
$$
{\bf \Phi} = {\eta{\bf r}\over r}\Big[{1\over \eta r}- \coth (\eta r)\Big]
\eqn\BPS
$$
from which we compute
$$
\big(\eta^2 - |{\bf \Phi}|^2\big)= {1\over r^2} \Big[ 2(\eta r)\coth (\eta
r) - (\eta r)^2{\rm cosech}^2(\eta r) -1\Big]\ ,
\eqn\BPSa
$$
and hence the function $U$. Note that we get an asymptotically flat solution
with the scalar field $\sigma$ tending to zero for all values of the
integration constant $\eta$ and hence for all values of the length of the Higgs
field
$\bf \Phi$ at infinity.

The construction of [\HL] can be be viewed [\GHL] (see also [\Kh]) as a
dimensional
reduction of a fivebrane solution of the field theory limit of the
ten-dimensional
heterotic string. This ten-dimensional supergravity/YM
theory can be reduced to five-dimensions and the resulting action can be
consistently truncated such that the only surviving fields are the 5-metric,
the dilaton, $\phi$, the two-form
potential, with 3-form field strength $H$, and the Lie-algebra valued
Yang-Mills
gauge potential, ${\bf Y}$, with two-form YM field strength ${\bf M}$. The
five-dimensional action for these fields is
$$
S= \int d^5x\; \sqrt{-g} e^{-2 \phi} \Bigl (R + 4 (\partial \phi)^2 - { 1
\over 3} H^2 - |{\bf M}|^2 \Bigr)\ ,
\eqn\aseven
$$
where the three-form $H$ satisfies the modified Bianchi identity
$$
\partial_{[A} H_{BCD]} = {3\over2} {\bf M}_{[AB}\cdot {\bf M}_{CD]}\ .
\eqn\aeight
$$
Note that \aseven\ is not the bosonic sector of a five-dimensional supergravity
because it lacks the five-dimensional Higgs field and an abelian gauge
potential
to partner the scalar $\phi$. It is however, a consistent truncation of five
dimensional supergravity coupled to both a YM supermultiplet and an abelian
vector multiplet, and in this context the coefficient $3\over 2$ in \aeight\ is
fixed by five-dimensional supersymmetry, as we shall explain later. The field
equations of \aseven\ can be solved with Kaluza-Klein type boundary conditions
to give the soliton solution of [\HL] of the dimensionally reduced
four-dimensional
field theory.

The purpose of this paper is to clarify the relation between the results of
[\HL]
and those of [\GKLTT]. Note first that these multi-monopoles exist for
{\it all} values of the dimensionless parameter $4 \pi G \eta ^2$ governing the
relative strength of (super)gravitational versus YM/Higgs forces. That is,
regardless of the ratio of the Higgs mass to the Planck mass, BPS monopoles do
not
undergo gravitational collapse to form black holes. In [\GHL], this feature was
attributed principally to the dilaton but since forces due to scalar fields are
{\sl
attractive} it seems unlikely that this is the explanation. For the solutions
of
[\GKLTT] this feature seems to be a consequence of the electrostatic
{\sl repulsion} brought about by the vector field, which also allows the
solutions
to saturate a Bogomolnyi-type energy bound. This interpretation is less clear
in the
context of the solutions of [\HL] because the string inspired five-dimensional
action \aseven\ has a two-form potential rather than a vector potential.
However, in
five dimensions a two-form potential can be exchanged for a vector potential by
a
duality transformation. This may be accomplished by imposing the constraint
\aeight\
by a Lagrange multiplier one-form potential $V$, and then promoting $H$ to the
status of an independent field (a procedure that is consistent with
supersymmetry
[\NT]). One introduces a new vector potential $V$ as a Lagrange multipler field
and
adds to the action \aseven\ the Lagrange multipler term
$$
S_L = {2\over3}\int d^5x\; V_E\, \epsilon ^{ EABCD} \Bigl ( \partial_A
H_{BCD} - {3\over2}  {\bf M}_{AB}\cdot {\bf M}_{CD} \Bigr )\ .
\eqn\anine
$$
Variation of the combined action with respect to $H_{ABC}$ reveals that
$$
H_{ABC}= {1\over 2} e^{2 \phi}~\epsilon _{ABCDE} F^{DE}\ ,
\eqn\aeleven
$$
where $F_5=dV$ is the two-form field strength of $V$.
One may now back substitute into the action \aseven\ augmented by
\anine\ to obtain the new, dual, action
$$
\tilde S = \int d^5x\; \Big\{ e^{-2 \phi}\sqrt{-g} \Big[ R + 4 (\partial
\phi)^2 -
F_5^2  - |{\bf M}|^2\Big] - V_A\, \epsilon^{ABCDE}{\bf
M}_{BC}\cdot {\bf M}_{DE} \Big\}
\eqn\atwelve
$$

Here we pause to remark that the unit coefficient of the last, topological,
term in
this action is determined by the $3/2$ coefficient in \aeight. It is also
precisely
what is required by supersymmetry. To see this, one needs to compare \atwelve\
with
the results of [\GST] for the coupling of five-dimensional supergravity to
vector
multiplets. To do this it is convenient to rescale the metric by $g_{AB}
\rightarrow
e^{{4 \over 3}\phi}~ g_{AB}$. Discarding a surface term, one then obtains the
dual
action in Einstein conformal gauge:
$$
\tilde S =\int d^5x \,\Big\{ \sqrt{-g}[R -{4\over 3}  (\partial \phi)^2 -
e^{{ 8 \over 3}\phi} ~F_5^2  -  e^{ -{ 4 \over 3} \phi }~ |{\bf M}|^2]
 - V_A\, \epsilon ^{ABCDE} {\bf M}_{BC}\cdot {\bf M}_{DE}\Big\} \ .
\eqn\afourteen
$$
By choosing the YM group to be $U(1)$ we can compare with the bosonic sector of
the
Maxwell/Einstein supergravity action which we review in an appendix. One finds
agreement provided that the coefficient of the topological term is as given
above.

We now have a form of the five-dimensional string-related action in which a
vector
potential replaces the two-form potential. To relate this to the action \aone\
we
must dimensionally reduce it to four-dimensions and then truncate to the fields
of
\aone. The dimensional reduction can be done by setting
$$
\eqalign{
(ds_5)^2 &= e^{2 \rho} (dx^5 - 2K)^2 + e^{ 2 \psi} (ds_4)^2\cr
V &= v\, (dx^5 - 2K) + A\cr
\phi &= \psi + { 1 \over 2} \rho \cr
{\bf Y} &= {\bf \Phi}\, (dx^5- 2K) + {\bf B} }
\eqn\bone
$$
where $\psi$, $\rho$, $v$ and ${\bf \Phi}$ are four-dimensional scalar fields,
$K$, $A$ and ${\bf B}$ are one-forms on four-dimensional spacetime and the
5-metric is in string conformal gauge. The particular choice of
four-dimensional
fields in \bone\ ensures that $A$ and ${\bf B}$ are invariant under the KK
gauge
transformation $K \rightarrow K + df$ induced by the coordinate transformation
$x^5\rightarrow x^5 + f(x^\mu)$. It is convenient to define the `modified'
four-dimensional field strength two-forms
$$
F'= F -2v dK\qquad ,\qquad {\bf G}' = {\bf G} -2{\bf \Phi}dK\ ,
\eqn\abfive
$$
where $F=dA$ and ${\bf G}$ is the four-dimensional YM field strength for ${\bf
B}$. On substitution of the ansatz \bone\ into the five-dimensional action
\atwelve\
one obtains, up to a surface term, the four-dimensional action
$$
\eqalign{
S = \int d^4x &\bigg\{ \sqrt{-g}\big[R- 2(\partial \psi)^2 -(\partial \rho) ^2
 -2 e ^{4\psi}(\partial v)^2  - e^{- 2 \psi + 2 \rho} (L)^2 \cr & - e^{ 2
\psi + 2 \rho} (F')^2 - 2 e^{-2 \rho} |{\cal D} {\bf \Phi}|^2
- e^{- 2 \psi} |{\bf G}'|^2  \big]\cr
&  - v\,  \epsilon ^{\mu \nu \lambda \rho } {\bf G}'_{\mu \nu}\cdot
{\bf G}'_{\lambda \rho} - 4A_\mu\, \epsilon ^{\mu \nu \lambda \rho }
{\bf G}'_{\nu\lambda}\cdot {\cal D}_\rho{\bf \Phi} \bigg\}\ , }
\eqn\bfive
$$
where $L_{\mu\nu} = 2\partial_{[\mu} K_{\nu]}$.

In order to obtain the action \aone\ by a truncation of \bfive\ we must choose
$$
a=\sqrt{3}
\eqn\vala
$$
and set
$$
v =0\ ,\qquad K=0\ , \qquad \rho= -{2\over \sqrt{3}}\sigma \ ,\qquad \psi=
-{1\over \sqrt{3}}\sigma\ .
\eqn\rel
$$
However, {\it this truncation is not a consistent one}, in the sense that
solutions of the equations of motion of the truncated theory are not
automatically
solutions of those of the untruncated theory but will be so only if the
untruncated
fields satisfy constraints. This is the principal complicating factor in
relating
the results of [\GKLTT] to those of [\HL]. These constraints are

$$
\eqalign{
0&=\varepsilon^{\mu\nu\rho\sigma}{\bf G}_{\mu\nu}\cdot {\bf G}_{\rho\sigma}\cr
0&= \partial_\mu\Big[ |{\bf \Phi}|^2
\varepsilon^{\mu\nu\rho\sigma}F_{\rho\sigma} +
2\sqrt{-g}\,e^{{2\over\sqrt{3}}\sigma} {\bf G}^{\mu\nu}\cdot {\bf \Phi}\Big]\cr
0&=2 e^{{2\over\sqrt{3}}\sigma}|{\cal
D}{\bf \Phi}|^2 -|G|^2\ .}
\eqn\rela
$$
It is straightforward to verify that the solution \atwo\ of the field equations
of
\aone\ satisfies these constraints. We deduce from this that \atwo\ is also a
solution of the field equations of the untruncated four-dimensional action
\bfive,
and hence of the field equations of the five dimensional action \atwelve. It
follows
that the latter field equations admit the solution
$$
\eqalign{
ds_5^2 &= -dt^2 + U[(dx^5)^2 + d{\bf x}^2] \cr
V &= {1\over 2U} dt\cr
e^{2\phi} &= U\cr
{\bf Y} &= {\bf \Phi}dx^5 + {\bf B}_i dx^i\ .}
\eqn\mono
$$
where ${\bf \Phi}$ and ${\bf B}_i$ solve the flat space Bogomolnyi equations.
This
is the solution used in [\HL]. We conclude that, for the special case of
dilaton
coupling $a=\sqrt{3}$, the multi-monopole solution of [\GKLTT] is equivalent to
that
of [\HL].

The five-dimensional interpretation of the monopole solutions enables a class
of
dyon solutions to be found by the method of boosting in the fifth dimension
[\M].
This changes the asymptotic value, $\eta$, of the length of the Higgs field
${\bf
\Phi}$ but, since $\eta$ was arbitrary, this problem can be simply overcome by
choosing the initial asymptotic value of $|{\bf \Phi}|$ to have some other
value,
$\eta'$, and then adjusting $eta'$ such that $|{\bf \Phi}|\rightarrow \eta$.
Thus,
we first make the replacement
$$
dx^5 \rightarrow \gamma (dx^5 + \beta dt)\qquad \ ,\qquad
dt \rightarrow \gamma (dt + \beta dx^5)\ ,
\eqn\ctwo
$$
in \mono, where $\gamma = (1-\beta ^2)^{ - {1 \over 2}}$. If we denote by ${\bf
\Phi}^0( {\bf x}, \eta') , {\bf B}^0_i ( {\bf x}, \eta')$, the solution of
the flat space Bogomolnyi equations ($\eta'$ being the expectation value
of the Higgs field that we start with) and by $U^0 ({\bf x}, \eta')$ the
associated solution of Poisson's equation, then the new fields are given by
$$
\eqalign{
ds_5^2 &= -{U^0 \over \gamma^2(U^0-\beta^2)} dt^2
+ \gamma ^2 (U^0 - \beta ^2 ) \big[ dx ^5
+ \beta { U^0 -1 \over U^0 - \beta ^2 } dt \big]^2
+ U^0 d {\bf x} ^2 \cr
V &= { \gamma\over 2 U^0 }\, dt +
{\beta \gamma \over 2U^0 }\, dx^5 \cr
e^{2 \phi} &= U^0 \cr
{\bf Y} &= \gamma {\bf \Phi}^0 ({\bf x}, \eta') dx ^5 + \beta \gamma {\bf
\Phi}^0 ({\bf x}, \eta')dt  + {\bf B}^0_i ( {\bf x}, \eta')d x^i \ .}
\eqn\newfields
$$
By comparison with \bone\ we can now read off all the four-dimensional fields
of
the dyon solution of the field equations of \bfive, except that we learn only
the
combination $e^{2\psi}ds_4^2$ of the scalar $\psi$ and the 4-metric. However,
since
$\psi=\phi -{1\over2}\rho$ and $\phi$ is boost invariant we can deduce the new
value of $\psi$ from that of $\rho$, and hence the new 4-metric. The result is
$$
\eqalign{
ds_4^2 &= - {1\over \gamma\sqrt{U^0-\beta^2}}dt^2 + \gamma \sqrt{U^0-\beta^2}\,
d{\bf x}^2 \cr
K &= -{1\over2}\beta {(U^0-1)\over (U^0-\beta^2)}\, dt\cr
e^{2\rho} &= \gamma^2(U^0-\beta^2)\cr
e^{2\psi} &= {U^0\over \gamma\sqrt{(U^0-\beta^2)}}\cr
v &= {\beta\gamma\over2U^0}\cr
A &= {1\over 2\gamma (U^0-\beta^2)} dt\cr}
\eqn\gravifields
$$
and
$$
\eqalign{
{\bf \Phi} &= \gamma {\bf \Phi} ^ 0 ( {\bf x}, \eta')\cr
{\bf B} &= \beta\gamma {\bf \Phi}^0( {\bf x}, \eta') dt + {\bf
B}_i^{(0)}({\bf x},\eta') dx^i dx^i\ .}
\eqn\dyon
$$
If we now set
$$
\eta'= \eta/\gamma\ ,
\eqn\etanew
$$
then the Higgs field is
$$
{\bf \Phi} =\gamma{\bf \Phi}^0({\bf x},\gamma^{-1}\eta)\ ,
\eqn\newHiggs
$$
which has the property that $|{\bf \Phi} | \rightarrow \eta$ as $|{\bf x} |
\rightarrow \infty$. We have thus arranged for the dyon to have the same
asymptotic
value for the Higgs field as it originally had for the monopole solution.
Observe
also that (for all values of $\eta '$) both $\rho$ and $\psi$ still vanish as
$|{\bf x}|\rightarrow \infty$, as they did in the monopole solution, and that
the
4-metric remains asymptotically flat. The $v$ field however is now non-zero at
spatial infinity, so the dyon is nevertheless not a solution in the same vacuum
as
that of the monopole. This feature seems to be the principal difference between
the
flat space case and its gravitational generalization.

We conclude with some remarks about the moduli space of the
antigravitating multi-monopole solutions discussed above. First, because the
solutions are constructed entirely in terms of a solution of the {\sl flat
space}
Bogomol'nyi equations the moduli space of these solutions is  {\sl
topologically}
the same as as for the flat space solutions, i.e. it is  diffeomorphic to the
space
of rational functions of a complex variable of degree $k$ where $k$ is the
monopole number. Let us now turn to the {\it metric} on this moduli space.
Because the monopole is a
solution of an $N=4$ supergravity theory that breaks half the supersymmetry,
the
metric should be hyper-Kahler. Moreover, it should be invariant under the
action of the Euclidean group. Hyper-Kahler metrics are rather rigid and given
the
boundary conditions and topology it is difficult to see how the metric can
differ
from the hyper-Kahler metric of the flat space theory. Consider, for example,
the
case of two monopoles. The metric on the `relative' moduli space is
four-dimensional and admits an $SO(3)$ action which rotates the complex
structures.
This fixes it to be the Atiyah-Hitchin metric.

If indeed the metric on the moduli space of $k$ BPS monopoles is the same as in
the flat space theory, it is presumably because the gravitational, gravivector
and
graviscalar interactions cancel against one another. In particular, this
cancellation must occur for large monopole separations, where it may easily be
checked. The lowest order two-body velocity dependent forces at large
separation
are given by the Darwin Lagrangian, which contains terms of the form [\GR]
$$
\eqalign {
&{{\bf v}_1 ^2 + {\bf v}_2 ^2  \over r_{12} } \,
\bigl ( 3 M_1 M_2 - \Sigma_1 \Sigma_2  \bigr )\cr
&{{\bf v}_1\cdot {\bf v}_2  \over r_{12} }\,
\bigl( Q_1 Q_2 + \Sigma_1 \Sigma_2 - 7 M_1 M_2  \bigr )\cr
&
{({\bf v}_1\cdot \hat{\bf r}) ({\bf v}_2 \cdot \hat {\bf r} )   \over r_{12}}
\,
\bigl( Q_1 Q_2 -\Sigma _1 \Sigma _2 -  M_1 M_2 \bigr )  }
\eqn\dtwo
$$
where $\Sigma_i$ are the scalar charges, $Q_i$ are the vector charges and $M_i$
are
the masses of the monopoles.  All three terms vanish if and only $a^2=3$ [\Ru]
which,
as we have shown earlier, is the value required to interpret the
four-dimensional BPS
monopoles as solutions of string theory. It is interesting to note that the
moduli
space of extreme electrically charged dilaton black holes is not only
asymptotically
flat (i.e. to order ${1\over r^2}$) when $a=\sqrt{3}$ but everywhere flat
[\Sh]. The
same is true of extreme magnetically charged $a=\sqrt{3}$ dilaton black holes
[\Ru] which can be viewed four-dimensional projections of Kaluza-Klein
monopoles
[\Sor,\GP]. In other words, the phenomenon of {\it enhanced} anti-gravity, i.e.
the
cancellation of gravitationally induced forces to first non-trivial order in
velocities, seems to be a general feature $a=\sqrt {3}$

If the metric on the moduli space is, as we suggest, unchanged by graviton,
gravi-photon and gravi-scalar exchange forces then the result of Sen [\Sen]
concerning a unique $L^2$ harmonic form on the relative moduli space
remains true in our case. Tensoring with the sixteen-plet of forms on the $S^1
\times {\R}^3$ factor (due to the centre of mass motion and the total electric
charge) will give a short Bogolmol'nyi $16$-fold supermultiplet of bound
monopole-dyon pairs.

Eigenfunctions of the Hodge-De-Rham Laplacian on the relative moduli space with
{\it non-vanishing} eigenvalues yield long, non-Bogolmolnyi, $256$-fold
supermultiplets of bound monople-dyon pairs. This follows from the fact that
the
non-vanishing eigenvalues come in multiples of sixteen, and hence give
256-plets on
tensoring with the centre of mass 16-plet. To see that the multiplicity of
non-zero eigenvalues is a multiple of sixteen it suffices to note, following
Hawking
and Pope [\HP], that the moduli space admits two covariantly constant chiral
spinors as a consequence of the fact that its holonomy lies in $Sp(1) \equiv
SU(2)$.
Using these spinors, Hawking and Pope show that the non-zero spectrum of the
the
Hodge-De-Rham Laplacian on $p$-forms, $p=1,2,3,4$ is given entirely in terms of
the
spectrum of the ordinary Laplacian on zero-forms, i.e. scalar functions.  For
each
scalar eigenfunction they showed that there are four eigen one-forms, six eigen
two-forms, four eigen three-forms and one eigen four-form, all with the same
eigenvalue. This implies a multiplicity of sixteen for all but the zero-mode
spectrum of the Hodge-De-Rham operator on the relative moduli space. Of course,
this
argument does not establish the existence of $L^2$ eigenforms with non-zero
eigenvalues, but merely that if they do exist then they must do so in multiples
of
sixteen.  However, the results of Gibbons and Manton [\GM] indicated strongly
that
$L^2$ scalar eigenfunctions exist on the relative moduli space and this
suggestion
has been confirmed by detailed calculations of Shroers and Manton [\SM].

\vskip 0.5cm

{\bf Appendix: Five-dimensional Einstein-Maxwell Theory}
\vskip 0.3cm

In this appendix we shall justify our claim that the coefficient of the
`topological' interaction term in \afourteen\ arises from the requirements of
supersymmetry.

The coupling of five-dimensional supergravity to $n$ vector multiplets
has been described in detail in [\GST]. The bosonic field content comprises the
metric, $g_{AB}$, $(n+1)$ vector fields $A_A^I$, $I=1, \dots ,n+1$ and $n$
scalar fields $\phi^i$, $ i=1,\dots,n$. If the gauge group is
abelian then the bosonic Lagrangian is
$$
R- g_{ij}(\phi) \partial \phi ^i \partial \phi ^j - { 1 \over 2} m_{IJ}(\phi)
F^I F^J + { 1 \over{ 3\sqrt 6}}\epsilon ^{ABCDE} A_A^I ~F_{BC}^J ~F _{DE}^J
{}~C_{IJK}
\ ,
\eqno (A.1)
$$
where $g_{ij}(\phi)$ is the metric on the scalar field target space,
$m_{IJ}(\phi)$ is a positive definite matrix-valued function of $\phi^i$, and
$C_{IJK}$ are {\sl constants}. These constants determine a symmetric
homogeneous
polynomial of degree three:
$$
{\cal N} ( \xi) = \beta ^3 C_{IJK} \xi ^I \xi ^J \xi ^K
\eqno (A.2)
$$
with $\beta = \sqrt{{2 \over 3}}$, where $\xi^I$ are the components of a
vector in an $n+1$ dimensional vector space $J$.
The scalar field target space is the ${\cal N}=1$ hypersurface in $J$. All
couplings of the theory are determined by ${\cal N}$.
For example
$$
m_{IJ} = - { 1 \over 2} \partial _I \partial _J \ln {\cal N}
\big |_{{\cal N}=1} \ .
\eqno (A.3)
$$
The target space metric $g_{ij}$  is given by
$$
g_{ij} = { 1 \over {\beta^2}} m_{IJ} h^I,_i h^J, _j \big |_{{\cal N}=1}
\eqno (A.4)
$$
where
$$
h^I= { 1 \over { 3 \beta }} m^{IJ} \partial _J {\cal N}\big |_{{\cal N}=1}\ .
\eqno (A.5)
$$

The pure five-dimensional supergravity corresponds to the case $n=0$. Let
$C_{111}=\gamma ^3 $. Then ${\cal N}= (\gamma\beta \xi)^3$ and thus the
hypersurface
${\cal N}=1$ is given by $\xi= (\beta \gamma)^{-1}$. There are no scalars and
just
one component of $m_{IJ}$, $m_{11}= \gamma ^2$. The resulting Lagrangian is
$$
R- { 1 \over 2} \gamma ^2 F^2  + { {\gamma^3}  \over { 3\sqrt 6}} \epsilon
^{ABCDE}
A_A F_{BC} F _{DE}\ .
\eqno (A.6)
$$
We recover the Lagrangian used in [\GKLTT] by setting $\gamma= \sqrt 2$.

In the case $n=1$ we set $C_{122}= {\gamma ^3 \over 3}$. Thus
$$
{\cal N}= {\beta \gamma}^3  \xi ^1 (\xi ^2)^2\ .
\eqno (A.7)
$$
The hypersurface ${\cal N}=1$ can be parametrised by the scalar $\phi$ by
setting
$$
\xi ^1 = (\beta \gamma )^{-1} \alpha ^2 e^{-{4\over3}\phi}
\eqno (A.8)
$$
and
$$
\xi ^2 = (\beta \gamma )^{-1} \alpha^{-1} e^{{2\over 3}\phi}\ ,
\eqno (A.9)
$$
where $\alpha$ is a constant. We find that the matrix $m_{IJ}$ is given by
$$
m= { (\beta \gamma )^2 \over { 2 \alpha ^4}}
\pmatrix{e^{{8\over 3}\phi}&0\cr 0& 2\alpha^6 e^{-{4\over 3}\phi}}\ ,
\eqno (A.10)
$$
and therefore
$$
h^I= \gamma^{-1} \alpha ^2 \bigl ( e^{-2\phi}, \alpha ^{-3} e^ {{2\over 3}\phi}
\bigr)\ ,
\eqno (A.11)
$$
whence $g_{ij} \partial \phi ^i \partial \phi ^j= {4\over3}(\partial \phi)^2$.
We may choose $\alpha ^6 =1$ in order to arrange that $m_{IJ}=\delta _{IJ}$ at
the
origin $\phi=0$, and we then choose $\gamma$ such that $(\beta \gamma)^3=-4$.
The
resulting lagrangian is
$$
\eqalign{
R - &{4\over 3}(\partial \phi)^2
- e^{{8\over3}\phi}~ F^{(1)}_{AE}~ F^{(1)}{}^{AE}
- e^{-{4\over3}\phi}~ F^{(2)}_{AE}~ F^{(2) }{}^{AE}\cr
&-  \epsilon ^{ABCDE} A ^{(1)} _E~ F^{(2)} _{AB}~F ^{(2)} _{CD}\ .}
\eqno (A.12)
$$
which is the five-dimensional string related action \afourteen\ for the special
case
of a $U(1)$ Yang Mills gauge group. As a further check we note that (A.6) is
recovered by the consistent truncation $\phi=0$, $A^{(2)} = \sqrt 2 A^{(1)}
=\sqrt{{2 \over 3}}\, { A}$.

\refout
\bye